\documentclass[twocolumn,floats,aps,amsmath]{revtex4} 
\usepackage{graphicx}
\usepackage{rotating}
\usepackage{dcolumn}
\usepackage{bm}
\usepackage{amssymb} 
\usepackage{amsmath} 
\usepackage[utf8]{inputenc} 

\begin{document}

\preprint{APS/123-QED}

\title{Superconductivity in Single-Crystal YIn$_3$}
\author{S.D. Johnson}
\affiliation{Physics Department, University of California Davis, Davis, CA 95616}
\author{J.R. Young}
\affiliation{Physics Department, University of California Davis, Davis, CA 95616}
\author{R.J. Zieve}
\affiliation{Physics Department, University of California Davis, Davis, CA 95616}
\author{J.C. Cooley}
\affiliation{
Los Alamos National Laboratory, Los Alamos, NM 87545}
\date{\today}

\begin{abstract}

We measure the superconducting transition of YIn$_3$ by resistivity,
susceptibility, and specific heat.  Despite using high-quality
single-crystal samples, the transitions detected by the three techniques
are shifted from each other in temperature, suggesting a region of
filamentary superconductivity.  We discuss the possible implications for
filamentary superconductivity in unconventional superconductors. 

\end{abstract}

\maketitle

\noindent

\section{Introduction}

One hundred years after the discovery of superconductivity, the goal of
predicting whether a material will superconduct from its atomic
composition and crystal structure remains elusive. Even more difficult is
estimating the transition temperature $T_c$. Over the decades, researchers
have often examined families of related materials, trying to quantify the
effects of replacing a single atomic constituent with a related atom.  One
such family, investigated 40 years ago, had the cubic AuCu$_3$ structure.
Much work followed the belief that $T_c$ directly correlated to the
valence electron concentration \cite{havinga1, matthias1}.  During this
time a large number of compounds with AuCu$_3$ structure were synthesized
and tested for superconductivity at ambient pressure, e.g. MIn$_3$ (M = Y,
La, Lu) at reported $T_c$ of 0.78 K, 0.71 K, and 0.24 K, respectively
\cite{havinga2}. 

Subsequently the discovery of superconductors that contain magnetic ions
in their crystal structures overturned the traditional view that magnetism
and superconductivity cannot coexist.  The field of potential
superconductors vastly broadened with the inclusion of magnetic ions, and
several new families have appeared since the early heavy-fermion and
high-temperature superconductors. In heavy-fermion materials, conduction
electrons acquire an effectively ``heavy" mass, often hundreds of times
greater than the bare electron mass \cite{steglich}, from  interaction
with the magnetic ions.  Furthermore, it appears that the
superconductivity is unconventional in that magnetic interactions actually
cause the electron pairing required for superconductivity, a role played
by phonons in conventional superconductors.

One of the heavy-fermion superconductors is CeIn$_3$, with $T_c$ peaking
at 0.2 K under 26.5 kbar of hydrostatic pressure \cite{walker}. Related
heavy-fermions CeMIn$_5$ (M = Co, Ir, Rh) and PuXGa$_5$ (X = Co, Rh) are
also superconducting with ambient-pressure $T_c$ reaching 18.5 K in
PuCoGa$_5$.  CeIn$_3$ has the AuCu$_3$ structure, which has inspired new
studies of the AuCu$_3$ family of materials. Normal state properties have
been measured in YIn$_3$ via de Haas-van Alphen experiments
\cite{pluzhnikov}, thermal conductivity \cite{mucha}, and heat capacity
\cite{hale}, leaving it well characterized and a promising candidate for
further exploration of the interplay between crystal structure and
magnetic superconductivity.

Here we present measurements on much higher-quality samples of YIn$_3$
than those previously explored below 1 K \cite{havinga2}.  We confirm the
earlier ac susceptibility measurements of the superconducting transition
and also provide resistivity and specific heat data.  We find an unusual
shift in the transition temperature using the different techniques, which
may be connected to behaviors in related heavy-fermion compounds.

\section{Sample Preparation}

YIn$_3$ crystallizes in a cubic AuCu$_3$ structure with lattice constant
4.593 \AA\ \cite{pluzhnikov}.  The single crystal samples were grown from
an excess Indium flux at a ratio of approximately 50:1 in an alumina
crucible with an integral alumina frit.  The crucible was sealed in a
quartz tube under argon, held at 1150$\circ$C for 72 hours and slow cooled
to 450$\circ$C. The 450$\circ$C crucible was removed from the furnace,
inverted, and placed in a centrifuge to spin off the excess flux.  After
cooling to room temperature crucible was broken out of the quartz tube and
the samples removed from the crucible.  The starting materials were
99.9999\% purity indium and 99.9\% purity yttrium.

The measurements were performed on a ${}^3$He/${}^4$He dilution
refrigerator, except that the ac susceptibility measurement on sample A
was done on a ${}^3$He sorbtion cryostat.  The sample was mounted in close
proximity to the cryostat thermometer.  For resistivity and susceptibility
we took data on both warming and cooling through the transition, with no
sign of hysteresis.  For the susceptibility and heat capacity measurements
we did not polish or clean excess indium from the sample. For the
resistivity measurement we cut down and polished all sides, both to
improve the sample geometry and to ensure a good connection between the
leads and the YIn$_3$.

\begin{figure}[tb]
\begin{center}
\scalebox{0.46}{\includegraphics{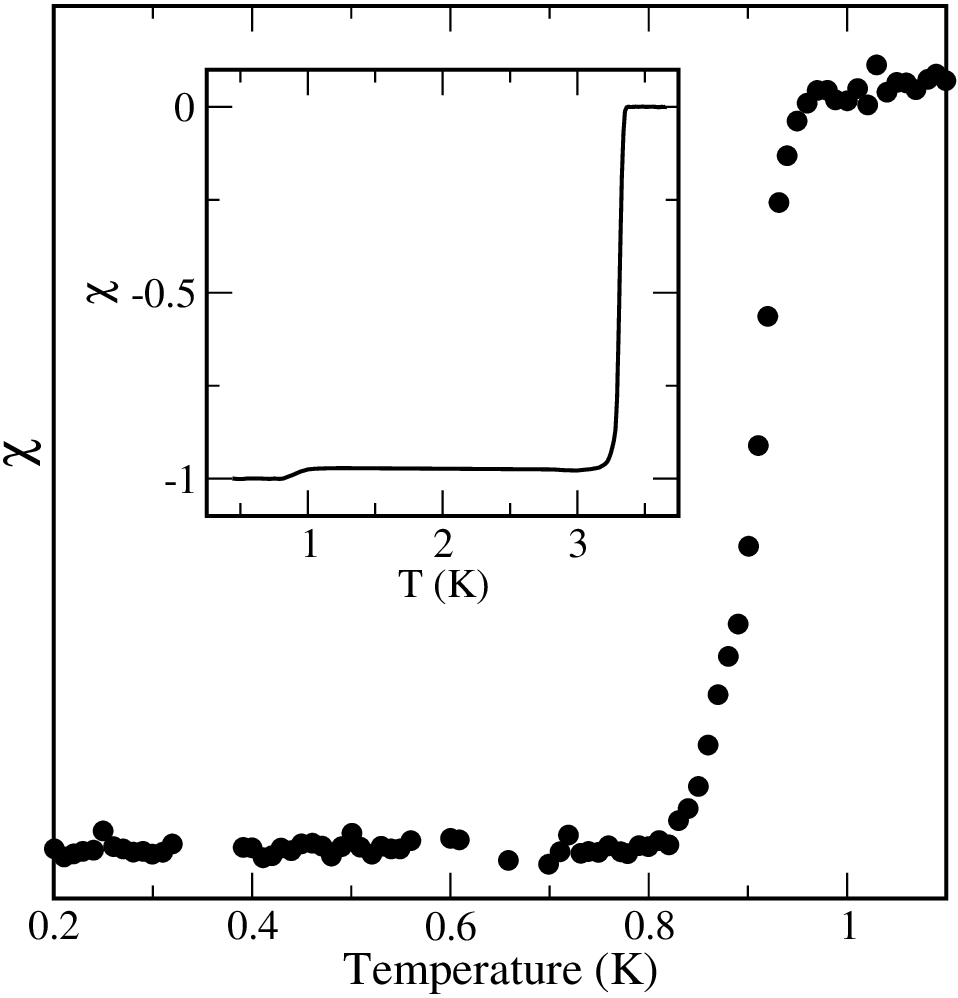}}
\caption{Susceptibility of sample B, with a sharp jump at 950 mK. The
inset shows susceptibility of sample A, measured to 3.65 K. The values are
scaled to 0 at the highest temperatures (sample fully normal) and -1 at
the lowest temperatures.  The large jump above 3 K indicates the
superconducting transition of the remnant indium in and around the
sample.  The contribution of YIn$_3$, barely visible near 1 K, makes up
only 3\% of the entire transition.}
\label{susc}
\end{center}
\end{figure}

\section{Results and Discussion}

We performed ac susceptibility measurements at a frequency of 12 Hz and an
ac field of about 0.4 gauss.  As shown in the main graph of Figure
\ref{susc}, the onset of the drop in susceptibility is at 960 mK, and the
sample is fully superconducting by 825 mK.  We scale the data to 0 in the
normal state and -1 in the superconducting state.  The inset shows
susceptibility extended to higher temperature. A much larger transition
appears at 3.4 K from the excess indium in and around the sample.  The
YIn$_3$ transition makes up only $\sim$3\% of the total susceptibility
change.  A simple estimate of the expected signal, based on the sizes of
the sample and the measurement coils and assuming the sample becomes
entirely superconducting, gives about 1.5 times the observed signal. This
reasonably good agreement suggests that in fact the indium is shielding
almost the entire sample volume. The small magnitude of the YIn$_3$
transition then stems from the shielding by the indium, and does not
indicate that only a small portion of the YIn$_3$ is superconducting.  We
believe the shielding arises from remnant indium on the surface of the
sample; subsequent cutting into the interior confirmed that only a few
small isolated veins of indium reside there.  

\begin{figure}[tb] 
\begin{center} 
\scalebox{0.4}{\includegraphics{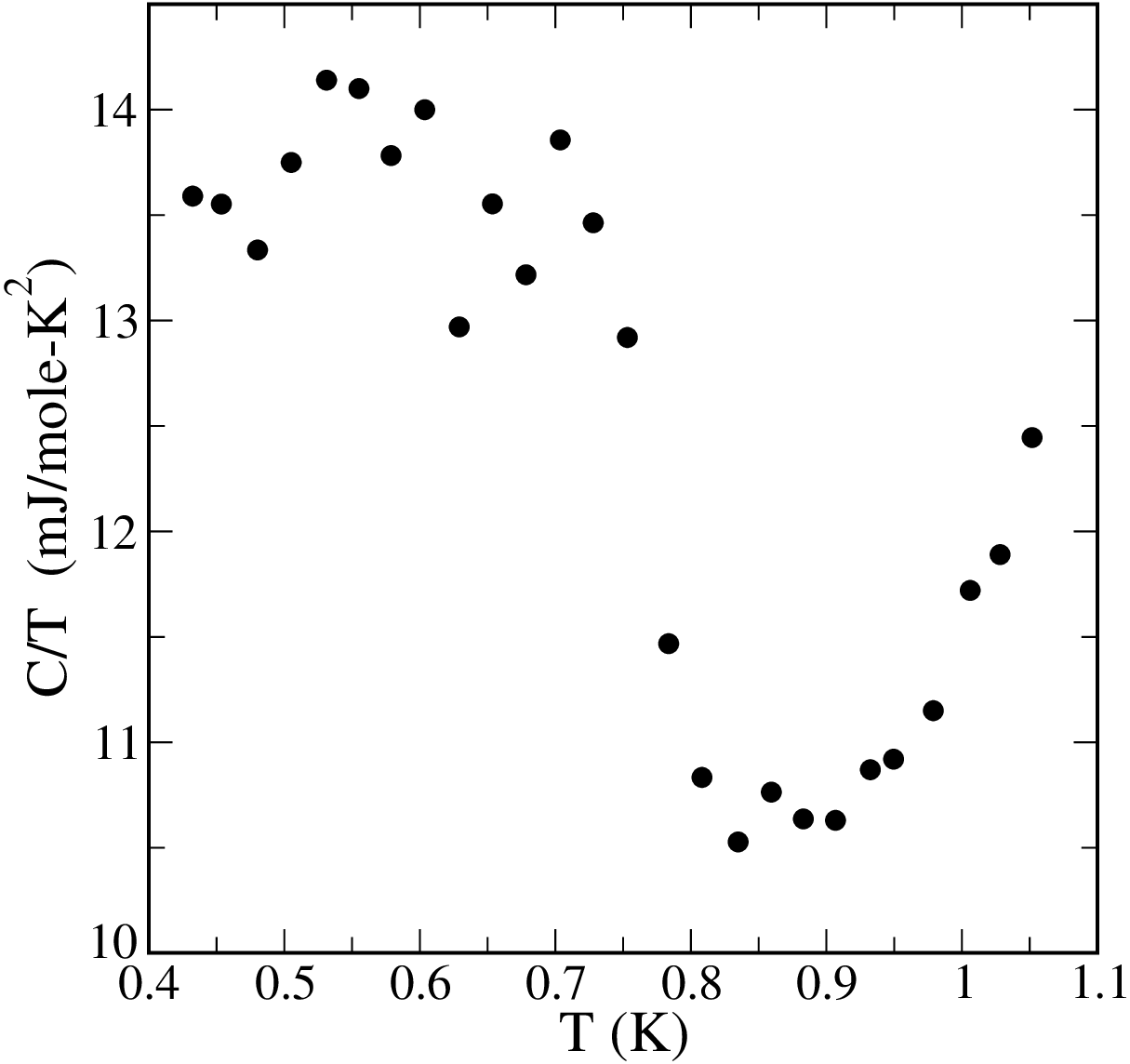}}
\caption{Heat capacity of YIn$_3$, sample B.} 
\label{HC} 
\end{center}
\end{figure}

Figure \ref{HC} shows the specific heat of sample B, with a feature at 825
mK.  We use a relaxation method with a 50:50 AuCr heater, a Cernox film
thermometer, and a graphite thermal link.  The sample mass is 57.55 mg,
and our heat pulses produce a temperature change of about 20 mK.  In the
normal state, $C/T$ is by no means constant, but it is the same order of
magnitude as in conventional superconductors.  This is consistent with de
Haas-van Alphen measurements reporting a small effective electron mass
\cite{umehara, pluzhnikov}.  The relative size of the transition is
$\frac{\Delta C}{C(T_c)}=0.35$.  The small size of the heat capacity
feature is another indication that the electrons which form pairs in the
superconducting phase have low effective mass. The jump size is probably
less than the BCS value of 1.43 in part because we made no adjustment for
any addendum heat capacity.

\begin{figure}[tb]
\begin{center}
\scalebox{.45}{\includegraphics{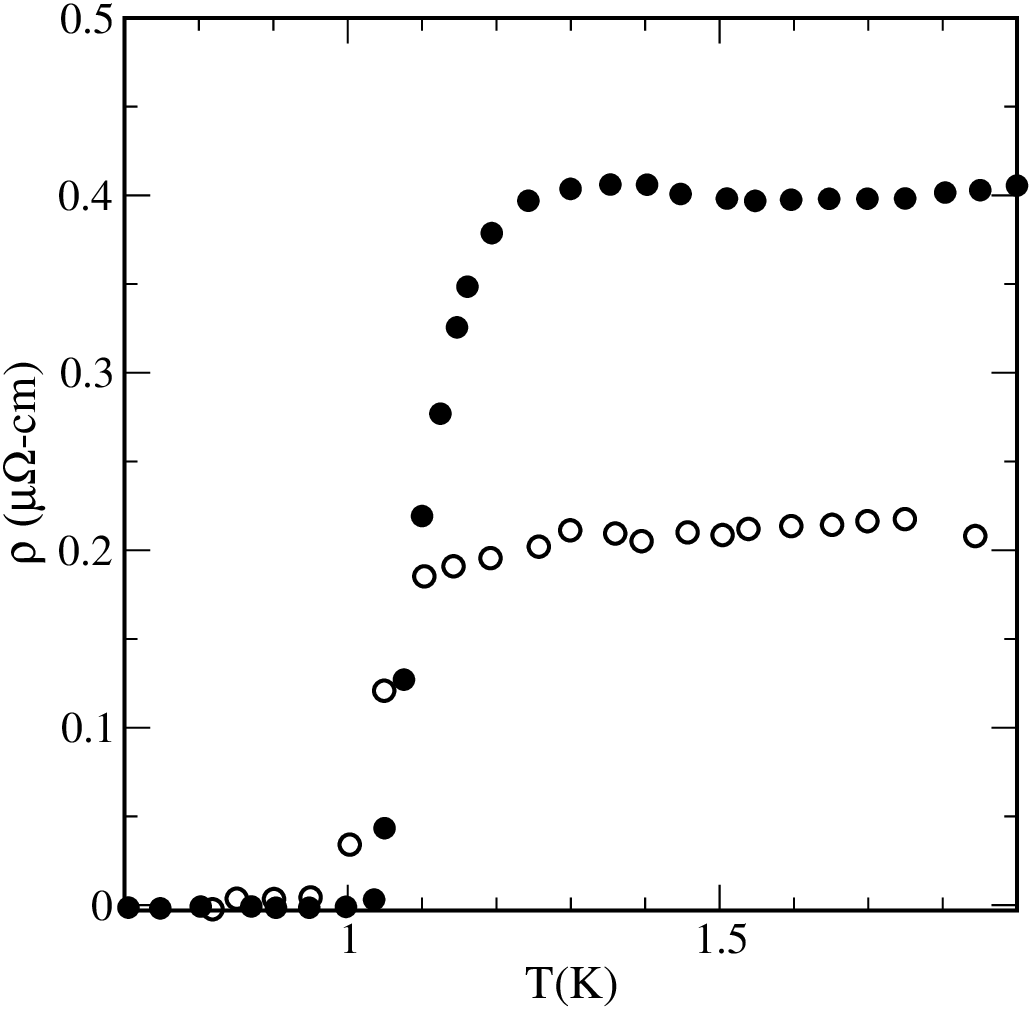}}
\caption{Resistive transitions for two samples of YIn$_3$.  Solid points are sample A; open are sample B.}
\label{resist}
\end{center}
\end{figure}

Figure \ref{resist} presents resistivity data for two YIn$_3$ samples. 
These were the same two samples shown in Figure \ref{susc}, but cut down
and polished to increase the aspect ratio and remove excess indium. We
used a four-wire technique with platinum wires 0.051 mm in diameter
spot-welded to the four corners.  The other ends of the wires were
soldered to a small copper plate which in turn connected to a resistance
bridge via fixed wiring on the cryostat. At room temperature the measured
resistivities of the two samples differ by about 10\%, with $\rho\approx
8\ \mu\Omega$-cm.  The resistivity ratios from room temperature to 4 K are
21 and 36 for samples A and B, respectively. A sharp drop in resistivity
with decreasing temperature indicates the superconducting transition. For
sample A the drop begins about 1.2 K and is completed above 1 K; for
sample B it occurs between 1.08 K and 0.98 K.

An earlier report \cite{havinga2} of superconductivity in YIn$_3$ was
based solely on ac susceptibility measurements.  The transition midpoint
was 780 mK, with half-width of 210 mK.  The half-width was calculated by
taking the tangent line to $\chi$ at the transition midpoint, and then
finding the temperatures where it attained the full normal and
superconducting susceptibilities.  Using this same definition, the
susceptibility transition shown in Figure \ref{susc} has half-width 36
mK.  Our sample has a higher transition temperature, a much smaller width,
and a fairly high residual resistivity ratio of 36.  All of these suggest
better sample quality than in the previous work. 

Strangely, measurements by resistivity, susceptibility, and heat capacity
give significantly different transition temperatures. For sample B, the
transition onsets are at 1.08 K, 0.95 K, and 0.90 K, respectively, with
the three techniques. The transition widths, from onset to completion, are
less than 100 mK in resistivity, and less than 150 mK with the other
techniques.  This means that the resistive transition does not overlap the
other two, and there is a gap of nearly 100 mK between the completion of
the resistive transition and the onset of the heat capacity transition. 
Although discrepancies in $T_c$ can result from sample quality issues, a
clear separation between the transitions such as we observe is unusual in
conventional superconductors.  However, it appears in a variety of
heavy-fermion and other unconventional superconductors, including
CeIrIn$_5$, URu$_2$Si$_2$, and Fe(Te,Se).  The discrepancy indicates a
regime of filamentary superconductivity, where despite a complete
transition in resistance the superconducting volume is too small to
register with bulk probes. 

In CeIrIn$_5$ the resistive $T_c$ occurs at 1.2 K, while the
susceptibility and heat capacity signals occur only at 0.4 K
\cite{petrovic}.  Improved samples have not brought the two closer
together; if anything, they have raised the resistive $T_c$. URu$_2$Si$_2$
also shows a large discrepancy in the superconducting phase boundary as
mapped out by resistivity or specific heat \cite{URu2Si2-phase}.  At
ambient pressure the values of $T_c$ are 1.4 K and 0.8 K, respectively. 
Both decrease with applied pressure, but the specific heat transition
disappears by 0.5 GPa, while the resistive transition persists to 1.8 GPa.
Similarly, at certain Se concentrations $x$, Fe$_{1+y}$(Te$_{1-x}$,Se$_x$)
exhibits a resistive transition without any specific heat signal.  The
onset of partial diamagnetism may \cite{FeTe-1.1} or may not
\cite{FeTe-gap} accompany the resistive transition.  These materials all
have different crystal structures, but they have in common a large and
highly anisotropic response to strain.  One possible source of the
filamentary superconductivity is this extreme sensitivity to uniaxial
pressure combined with internal strain within the sample \cite{bianchi}. 

One of these unconventional superconductors, CeIrIn$_5$, has a crystal
structure closely related to that of YIn$_3$.  As noted previously, its
parent compound CeIn$_3$ has the AuCu$_3$ lattice structure. In both these
Ce compounds, the magnetic cerium atom imparts a heavy effective mass to
the charge carriers.  Both become superconducting, and the sizes of their
specific heat discontinuities show that the heavy particles themselves
form the pairs necessary for superconductivity.  By contrast, with no
magnetic ions, YIn$_3$ has quasiparticle masses not far above that of free
electrons.  Its superconductivity is likely conventional, with electron
pairing mediated by phonons.  Our finding filamentary regime in a
conventional superconductor supports the internal strain explanation.
Further comparative studies of YIn$_3$, its cerium-based relatives, and
(Y,Ce)In$_3$ alloys may pinpoint how strongly the filamentary
superconductor is linked simply to the crystal structure, rather than to
any magnetic behaviors. 

\section{Conclusion}

We have measured ac susceptibility, specific heat, and
resistivity on two single-crystal samples of YIn$_{3}$ and confirmed
superconductivity by all three techniques.  We find a $T_c$ somewhat
higher from that previously reported and varying based on measurement
technique.  The temperature range between the resistive and bulk
transition temperatures supports filamentary superconductivity, which may
share the same origin as non-bulk superconductivity in several
unconventional superconductors. 

\section{Acknowledgements}
This work was funded by the NSF though grant
DMR-0454869 and the REU grant PHY-0649297.

\end{document}